\documentclass[prd,showpacs,nofootinbib,preprintnumbers]{revtex4}
\usepackage{amsmath} \usepackage{graphicx} \usepackage{amsfonts}
\usepackage{array} \usepackage{amsthm} \usepackage{bm}

\usepackage{latexsym}
\newcommand{\cQ}{{\cal Q}}\newcommand{\cP}{{\cal P}} \newcommand{\cM}{{\cal M}} \newcommand{\cN}{{\cal N}}

\newcommand{\e}{{\rm e}}

\newcommand{\lb}{\label}

\newcommand{\fr}{\frac}

\newcommand{\bw}{\begin{widetext}}
\newcommand{\ew}{\end{widetext}}
\newcommand{\be}{\begin{equation}}
\newcommand{\ee}{\end{equation}}
\newcommand{\bea}{\begin{eqnarray}}
\newcommand{\eea}{\end{eqnarray}}
\newcommand{\nn}{\nonumber}

\begin{document}
\begin{flushright}DTP-MSU/09-25
\end{flushright}

\title{Three-charge 2J black ring}

\author{Dmitri V. Gal'tsov} \email{galtsov@phys.msu.ru}
\affiliation{Department of Theoretical Physics, Moscow State
University, 119899, Moscow, Russia}

\author{Nikolai G. Scherbluk} \email{shcherbluck@mail.ru}
\affiliation{Department of Theoretical Physics, Moscow State
University, 119899, Moscow, Russia}

\date{\today}

\begin{abstract}
Using recently proposed new solution generating technique,
we construct the charged version of Pomeranski-Senkov doubly
rotating black ring in the $U(1)^3$ five-dimensional
supergravity. For arbitrary values of charges the solution is unbalanced,
but the Dirac-Misner string is removed when two of the charges are set to zero. In this particular case our solution can be uplifted to
some solution of six-dimensional vacuum gravity.

\end{abstract}

\pacs{04.20.Jb, 04.50.+h, 04.65.+e}

\maketitle

\section{Introduction}
Since the discovery of black rings~\cite{er} in five dimensions, many efforts were devoted to find more general configurations with the same topology of the horizon. The most simple black rings are   solutions of vacuum five-dimensional gravity. A considerable progress was achieved in the vacuum case due to application of  inverse scattering  method \cite {soliton}. Black rings were also generalized to configurations endowed with electric charges and magnetic dipole moments within  different models involving vector and scalar fields \cite{elva}. A particularly interesting model which has clear microscopic interpretation is based on supertubes \cite{MT}, and can be regarded as  M2-M5 intersections of 11D supergravity, or, equivalently, the D1-D5-P system of IIB theory. Within this theory a number of charged black rings were found \cite{3charge}, but neither of them contained two independent rotation parameters. Ultimately one is interested in the nine-parameter solution with independent mass, two rotation parameters, three charges and three dipole charges, and such task is still far from being completed.

In purely five-dimensional treatment this theory is equivalent to  $U(1)^3$ 5D supergravity with three vector and three constrained scalar fields.
Aiming to progress in obtaining the nine-parametric ring, we proposed a new solution generating technique based on dimensional reduction of $U(1)^3$ 5D supergravity to three dimensions where it is equivalent to gravity coupled sigma model on the coset $SO(4,4)/(SO(2,2)\times SO(2,2))$ or $SO(4,4)/(SO(4)\times SO(4))$ \cite{gtu1_3,imp}.  This approach generalized that of \cite{bccgsw,gc07}  devoted to minimal 5D supergravity in which case the coset is $G_{2(2)}/SL(2,R)^2$ (see \cite{cbjv} for further discussion and applications and \cite{mgun} for  more general mathematical aspects). In the case of black rings one can construct different realizations of these cosets performing dimensional reduction with respect to one angular direction or a linear combination of two independent rotation angles \cite{gisa}. In the second case the number of U-duality transformations preserving asymptotic flatness is larger, so one can hope to obtain a nine-parameter ring family starting with a seed with sufficient number of parameters.

It is tempting to start with Pomeranski-Senkov (PS) solution \cite{posen} for black ring with two rotation parameters obtained in the vacuum 5D gravity via soliton technique. As far as we are aware, only a few attempts to charge the PS black ring were undertaken \cite{bccgsw,hos}. In \cite{bccgsw} doubly rotating one-charge solution was constructed using G2 generating technique, but it contained the Dirac-Misner string. Later Hoskisson \cite{hos} derived two-charge solution using T-duality for a ten-dimensional uplifting of the PS solution. In this paper we  give the three-charge version of the PS solution which is a direct generalization of the one-charge solution of \cite{bccgsw}.  Eliminating two of three charges we get a very simple solution without the Dirac-Misner string. We show that this solution can be uplifted to some 6D vacuum solution, opening further perspectives of application of various generation methods.

Five-dimensional supergravity with three $U(1)$ vector fields can be derived from truncated toroidal compactification of    $11D$  supergravity:
 \be \label{ans11}
I_{11} = \frac{1}{16\pi G_{11}}\int\left(R_{11}\star_{11}
\mathbf{1}-\frac12 F_{[4]}\wedge \star_{11} F_{[4]} -
\frac16F_{[4]}\wedge F_{[4]} \wedge A_{[3]}\right),
 \ee where
$F_{[4]} = dA_{[3]} $, according to the ansatz
 \bea\label{met11}
ds_{11}^2 &=& ds^2_{5} + X^1 \left( dz_1^2 + dz_2^2 \right) + X^2
\left( dz_3^2 + dz_4^2 \right) + X^3 \left( dz_5^2 + dz_6^2
\right),
\\
A_{[3]} &=& \hat A^1 \wedge dz_1 \wedge dz_2 + \hat A^2 \wedge dz_3
\wedge dz_4 + \hat A^3 \wedge dz_5 \wedge dz_6 \,. \nn
 \eea
 Here $z_a=z^a,\
a=1,\ldots,6$ are the coordinates parameterizing the torus $T^6$.
The three scalar moduli $X^I,(I=1,2,3)$ and the three one-forms
$\hat A^I$ depend only on the five coordinates entering $ds^2_5$.
The moduli $X^I$ satisfy the constraint $X^1X^2X^3=1$,  implying
that the five-dimensional metric $ds_5^2$ is the Einstein-frame
metric. The reduced five-dimensional action reads: \bea\label{L5}
I_5 &=& \frac{1}{16 \pi G_5} \int \left( R_5 \star_5 \mathbf{1} -
\frac12 G_{IJ} dX^I \wedge \star_5 dX^J - \frac12G_{IJ} \hat F^I
\wedge \star_5 \hat F^J -
\frac{1}{6} \delta_{IJK} \hat F^I \wedge \hat F^J \wedge \hat A^K \right),\nn \\
G_{IJ}&=&{\rm diag}\left((X^1)^{-2},\ (X^2)^{-2},\
(X^3)^{-2}\right),\quad \hat F_{I}=d\hat A_{I},\quad I,J,K=1,2,3,\nn
 \eea
where the totally symmetric Chern-Simons coefficients
$\delta_{IJK}=1$ for the indices $ I,J,K $ being a permutation of 1,
2, 3, and zero otherwise.
\section{Generating technique}
Our generating technique is based on further reduction of
(\ref{met11}) to three dimensions assuming the 5D metric to depend only
on three coordinates. \bea
ds_5^2&=&\lambda_{pq}(dz^p+a^p)(dz^q+a^q)-\kappa\tau^{-1}h_{ij}dx^i
 dx^j,\quad \tau=-\det\lambda, \label{ds_5}\\
 \hat A^I &=&A^I_i(x^i)dx^i+\psi^I_pdz^p, \quad I=1,2,3,\quad
 p,q=7,8,
 \eea
where the three-dimensional metric $h_{ij}$ , the Kaluza-Klein (KK)
one-forms $a^p=a^p_idx^i$, and   the  moduli $\varphi_1,\varphi_2$ of $T^2$
 together with an axion $\chi$ entering the
symmetric $2\times 2$ matrix
 \be
\lambda_{pq}=\e^{-\frac{2}{\sqrt3}\varphi_1}\left(%
\begin{array}{cc}
  1 & \chi \\
  \chi & \chi^2+\kappa\e^{\sqrt3\varphi_1-\varphi_2} \\
\end{array}
\right)
 \ee
are independent on $z^p$. A factor $\kappa=\pm 1$ in
(\ref{ds_5}) is responsible for the signature of the direction
$z^8$: $\kappa=1$ for space-like, and $\kappa=-1$ for time-like. The
$5D$ $U(1)$ one-forms $\hat A^I$ reduce to the $3D$ one-forms
$A^I(x^i)$ and six axions collectively denoted as a 2D-covariant
doublet $\psi_p^I=(u^I,v^I)$ with the index $p$ relative to the
metric $\lambda_{pq}$.

To obtain the three-dimensional sigma-model one has to dualise the
one-forms $A^I$ and  $a^p$ to
scalars, which will be denoted as  $\mu_I$ and $\omega_p$ respectively:
 \bea
  && \tau\lambda_{pq}da^q=\star V_p, \quad V_p=d \omega_p- \psi_p^I  \Big( d \mu_I+\frac16 \delta_{IJK}d\psi_q^J
 \psi_r^K\varepsilon^{qr}\Big),\nn\\
 && dA^I=d\psi_q^I\wedge a^q+\tau^{-1}G^{IJ}\star
  G_J,\quad G_I=d\mu_I+\frac12\delta_{IJK}d\psi_p^J
 \psi_q^K\varepsilon^{pq}.\label{eqs_of_dual}
 \eea
This leads to the  gravity coupled $3D$ sigma-model
 \be
I_3=\frac{1}{16\pi G_3 }\int \sqrt{|h|}\left(R_3-{\cal
G}_{AB}\frac{\partial\Phi^A}{\partial
x^i}\frac{\partial\Phi^B}{\partial x^j}h^{ij}\right)d^3x,\label{L3}
 \ee
where the Ricci scalar $R_3$ is build using the 3-dimensional metric
$h_{ij}$. The set of the potentials
$\Phi^A=(\varphi_1,\varphi_2,X^1,X^2, \psi^I, \mu_I,\chi,\omega_p),$
$A=1,\ldots,16$ realizes the harmonic map
 between the 3D space-time  and the target
space (TS) with the metric ${\cal
G}_{AB}(\Phi^C)$:
 \bea
   dl^2 = {\cal{G}}_{AB}d\Phi^Ad\Phi^B&\equiv&\frac12 G_{IJ}(dX^IdX^J+d{{\psi^I}^T}\lambda^{-1}
d\psi^J)-\frac12\tau^{-1}G^{IJ}G_IG_J+\nn \\ &&+\frac14 \mathrm{Tr} \left(
\lambda^{-1} d\lambda \lambda^{-1} d\lambda \right)
  +  \frac14\tau^{-2} d\tau^2 - \frac12\tau^{-1} V^T \lambda^{-1}
V.\label{TS_metric}
 \eea
This target space is identified as $SO(4,4)/(SO(2,2)\times SO(2,2))$
for $\kappa=-1$ and  $SO(4,4)/(SO(4)\times SO(4))$ for
$\kappa=1$ with the isometry group $SO(4,4)$ acting transitively.
The action of isometries is  simplified if one introduces the
matrix representative $\cM$ of the coset which can be conveniently
chosen as the symmetric $8\times 8$  matrix:
 \be
 dl^2=-\frac18 \mathrm{Tr}(d{\cal M}d{\cal M}^{-1}).\label{TS_by_M}
 \ee
The $SO(4,4)$ isometry  transformations then will read
 \be
 {\cal M}\to {\cal M}'=g^T {\cal M}g,\label{trans_M}
 \ee
where $g$ is some  $SO(4,4)$ matrix.

The matrix  ${\cal M}$ can be written in terms of $4\times 4$
matrices $\cP, \; \cQ$ and $\widetilde{\cal P}=({\cal
 P}^{-1})^{\widehat T}$, where $\widehat T$ denotes transposition with respect to the minor diagonal:
 \be
 {\cal M}=\left( \begin {array}{cc}
 {\cal P}&{\cal P}{\cal Q}\\
 {\cal Q}^T{\cal P}&\widetilde{\cal P}+{\cal Q}^T{\cal P}{\cal Q}
 \end {array} \right),\quad \det{\cal M}=1.
 \ee
 The matrix $\cQ$, antisymmetric under $\widehat T$, reads
 \be
 {\cal Q}=\left( \begin{array}{cccc}
 \tilde{\mu}_1 &\tilde{\omega}_7
  &
 \tilde{\omega}_8 &0\\
  -v^2&-\tilde{\mu}_3 &0&\\
   -u^2&0&\cdots&\\
    0&&&\\
 \end{array} \right),
 \ee
where $\tilde{\mu}_I=\mu_I+w_I,\;\; w_1=u^{[3}v^{2]},\;\;
w_2=u^{[1}v^{3]},\;\; w_3=u^{[1}v^{2]}$ (brackets denote alternation
over indices with 1/2), and
$$\tilde{\omega}_7=\omega_7+\fr13(u^1w_1+u^3w_3) -\mu_2u^2,\quad
\tilde{\omega}_8=\omega_8+\fr13(v^1w_1+v^3w_3) -\mu_2v^2.$$ The
matrix $\cP$ then can be written in $3\times 3$ and $1\times 3$
blocks
 \be
 {\cal P}=\left( \begin {array}{cc}
 \Psi^T\Lambda \Psi,&\Psi^T\Lambda\Phi\\
 \Phi^T\Lambda\Psi,&\Phi^T\Lambda\Phi+\frac{X^1}{X^2}
 \end {array} \right),\quad\label{def_P}
 \ee
where
 $$
 \Psi=\left( \begin {array}{ccc}
 1&u^3&-v^3\\
 0&1&0 \\
 0&0&1 \end {array}
 \right),\quad
 \Lambda=\frac{X^3}{\tau}\left( \begin {array}{ccc}
 -(X^3)^{-1}&0&0\\
 0&-\lambda_{77}&\lambda_{78} \\
 0&\lambda_{78}&-\lambda_{88} \end {array}
 \right),\quad
 \Phi=\left( \begin {array}{ccc}
 \tilde{\mu}_2 \\
 -v^1\\
 -u^1\end {array}
 \right).
  $$

Since the target space variables are related to initial metric and
matter field via differential dualisation equations
(\ref{eqs_of_dual}), extraction of a new solution from the
transformed potential presents certain technical difficulties. To
avoid them one can transform in parallel the dualised matrix, which
is introduced as follows \cite{imp}.
  Introducing
the matrix-valued current one-form  ${\cal J}$
$$
{\cal J}={\cal J}_idx^i={\cal M}d{\cal M}^{-1}
$$
we can rewrite the 3-dimensional sigma-model action (\ref{L3}) as
$$
I_3=\frac{1}{16\pi G_3}\int \left(R_3\star 1-\frac18
\mathrm{Tr}({\cal J}\wedge\star{\cal J})\right).
$$
In this expression the Hodge dual $\star\ $ with respect
to the 3-dimensional metric $h_{ij}$ is introduced. Variation of this action with
respect to ${\cal J}$ shows that the two-form $\star{\cal J}$ is
closed: \be d\star{\cal J}=0.\label{eqs_of_motion}\ee Variation with
respect to the metric leads to  three-dimensional Einstein
equations: \be (R_3)_{ij}=\frac18 \mathrm{Tr}({\cal J}_i{\cal
 J}_j). \ee
 The first equation (\ref{eqs_of_motion}) means that the
 matrix-valued
 two-forms $\star {\cal
 J}$ is locally exact, i.e., it can be  presented
 as the exterior derivative of some matrix-valued one-form ${\cal
 N}$, that is
 \be
  \star{\cal J}={\cal M}\star d{\cal M}^{-1}=d{\cal N}.\label{def_N}
 \ee
 The matrix ${\cal N}$ is defined up to  adding an arbitrary matrix-valued
  closed one-form, which can be determined by choosing suitable
 asymptotic conditions. Now comparing the matrix dualisation equation
 (\ref{def_N}) with the initial dualisation equations
 (\ref{eqs_of_dual})
we find  the following purely algebraic relations between certain
 components of the matrix ${\cal N}^{ab},\ a,b=1,\ldots, 8$ are
and the previous variables $a^p$ and
  $A^I$, namely
 \bea
  && a^7={\cal N}^{16},\quad a^8={\cal N}^{17},\nn\\
  && A^{1}=\psi^1_{p}a^p+{\cal N}^{15},\quad A^{2}=\psi^2_{p}a^p+{\cal
  N}^{14},\quad A^{3}=\psi^3_{p}a^p-{\cal
  N}^{26}.\label{rel_for_N}
 \eea
 Therefore the metric and
 matter fields  can be extracted algebraically from the matrix ${\cal N}$.
Now, from the definition (\ref{def_N}) and
 the transformation law for the matrix ${\cal M}$ (\ref{trans_M})
 under the global transformations $g\in SO(4,4)$ is clear that
  the matrix ${\cal N}$ obey the tramsformation law
 $$
  {\cal N}\rightarrow {\cal N}\ '=g^T{\cal N}(g^T)^{-1}.
 $$
To obtain the matrix $\cN$ corresponding to the seed solution one has to solve the dualisation equations which is presumably a simple problem. Then one performs transformation of $\cN$ independently of the transformation of $\cM$. This allows to avoid the back dualisation of the transformed sigma-model variables to get the desired metric and matter fields.


\section{Sigma-model representation of Pomeransky-Senkov solution}
Using the inverse scattering technique Pomeransky and Senkov \cite{posen} derived the following vacuum black ring solution with two parameters of rotation :
\begin{eqnarray}
&&ds^2 = - \frac{H(y,x)}{H(x,y)} (dt + \Omega)^2 -
\frac{F(x,y)}{H(y,x)} d\phi^2 - 2\frac{J(x,y)}{H(y,x)} d\phi d\psi
\nonumber\\&&+ \frac{F(y,x)}{H(y,x)} d\psi^2 + \frac{2 k^2
H(x,y)}{(x-y)^2 (1-\nu)^2} \left[ \frac{dx^2}{G(x)} -
\frac{dy^2}{G(y)} \right],\label{PomSen}
\end{eqnarray}
where the coordinates ($t, x, y, \phi, \psi$) vary in the range
$-\infty < t < +\infty, -1 \leq x \leq 1, -\infty < y <-1, 0 \leq
(\phi, \psi) < 2 \pi,\;k,\,\nu,\,\lambda$ are parameters, and the
rotation one-form is   $\Omega = \Omega_\psi d\psi +
\Omega_\phi d\phi$. Explicitly,
 \bea \label{omega}
\Omega  &=& - \frac{2 k \lambda \sqrt{(1 + \nu )^2 -
\lambda^2}}{H(y,x)} \Bigl[ (1 - x^2) y \sqrt{\nu} d\psi
\nn\\
&+& \frac{1 + y}{1 - \lambda + \nu} [ 1 + \lambda - \nu + x^2 y \nu
(1 - \lambda - \nu) + 2 \nu x (1 - y) ] \, d\phi \bigr],\nn
 \eea
and the functions $G, H, J, F$ are  (we use here the original
notation of \cite{posen}, not to be confused with our quantities
$\lambda_{pq},x,y$):
\begin{eqnarray} \label{functions}
G(x) &=& (1 - x^2) (1 + \lambda x + \nu x^2),
\nonumber\\
H(x,y) &=& 1 + \lambda^2 - \nu^2 + 2 \lambda \nu (1 - x^2) y + 2 x
\lambda (1 - y^2 \nu^2) + x^2 y^2 \nu (1 - \lambda^2 - \nu^2),
\nonumber\\
J(x,y) &=& \frac{2 k^2 (1 - x^2) (1 - y^2) \lambda \sqrt{\nu}}{(x -
y) (1 - \nu)^2} \left[ 1 + \lambda^2 - \nu^2 + 2 (x + y) \lambda \nu
- x y \nu (1 - \lambda^2 - \nu^2) \right],\nn
\\
F(x,y) &=& \frac{2 k^2}{(x - y)^2 (1 - \nu)^2} \Bigl\{ G(x) ( 1 -
y^2) \Bigl( [ (1 - \nu)^2 - \lambda^2 ] (1 + \nu) + y \lambda (1 -
\lambda^2 + 2 \nu - 3 \nu^2) \Bigr) \nn\\
&+& G(y) \Bigl[ 2 \lambda^2 + x \lambda [ (1 - \nu)^2 + \lambda^2 ]
+ x^2 [ (1 - \nu )^2 - \lambda^2 ] (1 + \nu) + x^3 \lambda (1 -
\lambda^2 - 3 \nu^2 + 2 \nu^3) \nn\\
&-& x^4 (1 - \nu) \nu (-1 + \lambda^2 + \nu^2) \Bigr] \Bigr\}.
\nonumber
\end{eqnarray}
Regularity of the black ring implies the inequalities
$0\leq\nu<1,\;2\sqrt{\nu}\leq\lambda<1+\nu.$ The mass and angular
momenta can be read out from the asymptotic expansion of the metric:
$$
M = \frac{3 k^2 \pi \lambda}{G_5 (1 - \lambda + \nu )}, \quad J_\psi
= \frac{4 k^3 \pi \lambda \sqrt{\nu} \sqrt{(1 + \nu)^2 -
\lambda^2}}{G_5 (1 - \nu)^2 (1 - \lambda + \nu)},
$$
$$ \quad J_\phi
= \frac{2 k^3 \pi \lambda (1 + \lambda - 6 \nu + \lambda \nu +
\nu^2) \sqrt{(1 + \nu)^2 - \lambda^2}}{G_5 (1 - \nu)^2 (1 - \lambda
+ \nu)^2},
$$
where $G_5$ is the $5D$ Newton constant. This solution is free of
conical and Dirac string singularities.

This solution depends only on two coordinates $x,y$, and therefore possess three commuting Killing symmetries $\partial_t,\,\partial_\phi,\partial_\psi$. The charging transformation arise when one choose $z^8=t$ and $z^7$ is either one of the angular coordinates $\phi,\, \psi$, or some linear combination of the two \cite{gisa}. The last option is technically more complicated, though it open a way to get a reacher set of isometry transformations preserving asymptotic conditions relevant to black rings. Here we restrict to the case $z^7=\psi$. In this case the three-dimensional metric which remains unaffected by transformations reads:
\be h_{ij}dx^idx^j =
\frac{2 k^2}{(1-\nu)^2(x-y)^2} \left[F(y,x)
\left(\frac{dx^2}{G(x)}-\frac{dy^2}{G(y)}\right)-
\frac{2k^2G(x)G(y)} {(x-y)^2}d\phi^2 \right].
\ee

The target space potentials corresponding to dimensional reduction
of the metric (\ref{PomSen}) from $D=5$ to $D=3$ with respect to
$(t,\psi)$ are
\begin{eqnarray}\lb{seedpot}
\lambda_{88} &=& - \frac{H(y,x)}{H(x,y)}, \quad \lambda_{78} = -
\frac{H(y,x)}{H(x,y)} \Omega_\psi, \quad \lambda_{77} =
\frac{F(y,x)}{H(y,x)} - \frac{H(y,x)}{H(x,y)} \Omega_\psi^2, \quad
\tau = \frac{F(y,x)}{H(x,y)},\nn
\\
a^7_\phi &=& - \frac{J(x,y)}{F(y,x)},\qquad a^8_\phi = \Omega_\phi -
a^7_\phi \Omega_\psi ,\quad \hat A^I=0,\quad X^I=1.\nn
\end{eqnarray}
The corresponding block matrices forming the coset matrix $\cM$ will read
 \be
 {\cal P}=\left(%
\begin{array}{cc}
  \Lambda & 0 \\
  0 & 1 \\
\end{array}%
\right),\quad
\widetilde{\cal P}=\left(%
\begin{array}{cc}
  1 & 0 \\
  0 & \Lambda^{-1} \\
\end{array}%
\right),\quad
\Lambda=\tau^{-1}\left(%
\begin{array}{ccc}
  -1 & 0 & 0 \\
  0 & -\lambda_{77} & \lambda_{78} \\
  0 & \lambda_{78} & -\lambda_{88} \\
\end{array}%
\right),\quad
{\cal Q}=\left(%
\begin{array}{cccc}
  0 & \omega_7 & \omega_8 & 0 \\
  0 & 0 & 0 & -\omega_8 \\
  0 & 0 & 0 & -\omega_7 \\
  0 & 0 & 0 & 0 \\
\end{array}%
\right),
 \ee
$\Psi=\mathrm{diag}(1,1,1)$, and the column matrix $\Phi$ is zero.

We have to calculate also the dualised quantities $\omega_p, \,\mu_I$ entering the matrix $\cN$.
For our particular application we will need only the
quantities $\omega_8$ and some components of the matrix one-form ${\cal N}_i dx^i$. The first one is
the solution of the dualisation equation
$$
 \tau\lambda_{pq}h^{im}h^{jn}(\partial_m a_n^q-\partial_n
 a_m^q)=\frac{1}{\sqrt{h}}\epsilon^{ijk}\partial_k\omega_p,
$$
whose the $p=8$ component can be explicited  as (assuming
$\epsilon^{\phi x y}=1$, and denoting $h\equiv \det h_{ij}$):
 \bea
&&\partial_x\omega_8=\tau\sqrt{h}h^{\phi\phi}h^{yy}(\lambda_{78}\partial_y
a_{\phi}^7+\lambda_{88}\partial_y  a_{\phi}^8),\nn\\
&&
\partial_y\omega_8=-\tau\sqrt{h}h^{\phi\phi}h^{xx}(\lambda_{78}\partial_x
a_{\phi}^7+\lambda_{88}\partial_x  a_{\phi}^8).\nn
 \eea
The computation gives: $\omega_8=-\tilde\Omega_{\phi}$.

The following components of $\cN$ can be readily found
 \be {\cal N}^{16}=a^7=-\frac{J(x,y)}{F(y,x)}d\phi,\quad
{\cal
N}^{17}=a^8=\left(\Omega_{\phi}+\frac{J(x,y)}{F(y,x)}\Omega_{\psi}\right)d\phi.
 \ee
 Also, the absence of the electromagnetic potentials gives
${\cal N}^{15}={\cal N}^{14}={\cal N}^{26}=0$.
From non-trivial components, we will need only  ${\cal N}^{32}$, which
satisfies the following equations
 \bea
&& \partial_x {\cal
N}_{\phi}^{32}=-\frac{1}{\sqrt{h}}h_{xx}h_{\phi\phi}({\cal
M}\partial_y {\cal M}^{-1})^{32}=-\tilde\Omega_{\phi}\partial_y
a_{\phi}^7
+\frac{\lambda_{88}^2}{\tau}\frac{h_{xx}h_{\phi\phi}}{\sqrt{h}}\partial_x\Omega_{\psi},\nn\\
&& \partial_y {\cal
N}_{\phi}^{32}=\frac{1}{\sqrt{h}}h_{yy}h_{\phi\phi}({\cal
M}\partial_x {\cal M}^{-1})^{32}=\tilde\Omega_{\phi}\partial_x
a_{\phi}^7
-\frac{\lambda_{88}^2}{\tau}\frac{h_{yy}h_{\phi\phi}}{\sqrt{h}}\partial_y\Omega_{\psi}.\nn
 \eea
The solution can be presented as
 \be {\cal N}_{\phi}^{32}=-\tilde
a_{\psi}^8=-\tilde\Omega_{\psi}+\tilde
a_{\psi}^7\tilde\Omega_{\phi},
 \ee
where the quantities $a^7_{\psi}$ and $a^8_{\psi}$ are defined as
 \be
 a^7_{\psi}\equiv\frac{J(x,y)}{F(x,y)}=-\frac{J(y,x)}{F(x,y)}
 =a^7_{\phi}(y,x)=\tilde
 a^7_{\phi}(x,y),\quad
 a^8_{\psi}\equiv\Omega_{\psi}-\Omega_{\phi}a^7_{\psi},
 \ee
and the tilded functions of $x,y$ are introduced which correspond to
interchanging the arguments: $\tilde{f}(x,y)=f(y,x)$.
The remaining component of $\cN$ are more complicated to find, but we will not need them in what follows.

\section{The new solution}
We are interested by asymptotically flat solutions, so we have to select those elements of the $SO(4,4)$ isometry group which preserve asymptotic conditions under
$$
 {\cal M}'=g {\cal M}g,\quad {\cal N}\ '=g{\cal N}g^{-1}.
$$
where $g$ is a symmetric $8\times 8$ matrix. An analysis
\cite{gtu1_3} shows that within the reduction  along $t,\,\psi$
which was adopted here, such a property is fulfilled only for the
three-parametric subgroup given by the matrix \be
 g=
 \left( \begin {array}{cccccccc} c_3&0&-s_3&0&0&0&0&0\\0&c_1c_2&0&-
 s_1c_2&-c_1s_2&0&-s_1s_2&0\\
 -s_3&0&c_3&0&0&0&0&0\\0&-s_1c_2&0&c_1c_2&s_1s_2&0&c_1s_2&0
 \\0&-c_1s_2&0&s_1s_2&c_1c_2&0&s_1c_2&0\\0&0&0&0&0&c_3&0&s_3\\
 0&-s_1s_2&0&c_1s_2&s_1c_2&0&c_1c_2&0\\0&0&0&0&0&s_3&0&c_3\end {array}
 \right),\quad c_I\equiv \cosh(\alpha_I),\quad  s_I\equiv
 \sinh(\alpha_I),
\ee where $\alpha_1,\,\alpha_2,\alpha_3$ are real parameters. The
corresponding manipulations with matrices were performed using
computer. Having obtained the matrices ${\cal M}'$ and ${\cal N}\
'$, the extraction of the target space variables of the new solution
is straightforward. These are listed as
 \bea
&&\lambda_{77}'=D^{-2/3}\frac{(c\lambda_{78}+s\lambda_{88}\omega_8)^2}{\lambda_{88}}-D^{1/3}\frac{\tau}{\lambda_{88}},
\quad
\lambda_{78}'=D^{-2/3}(c\lambda_{78}+s\lambda_{88}\omega_8),\quad
\lambda_{88}'=D^{-2/3}\lambda_{88},\quad \tau'=D^{-1/3}\tau,\nn\\
&& a_{\phi}^7{}'=a_{\phi}^7,\quad a_{\phi}^8{}'=ca_{\phi}^8+s{\cal N}_{\phi}^{32},\quad
X^I{}'=\frac{D^{1/3}}{D_I},\quad c\equiv c_1c_2c_3,\quad s\equiv s_1s_2s_3, \nn\\
&& \hat A_t^I{}'=v^I{}'=s_Ic_ID_I^{-1}(1+\lambda_{88}),\quad
\hat A_{\psi}^I{}'=u^I{}'=D_I^{-1}(\lambda_{78}s_I\sum_{J<K}\delta^{IJK}c_Jc_K-\omega_8c_I\sum_{J<K}\delta^{IJK}s_Js_K),\nn\\
&&
A_{\phi}^I{}'=u^I{}'a_{\phi}^7+v^I{}'a_{\phi}^8{}'-s_I\sum_{J<K}\delta^{IJK}c_Jc_K
a_{\phi}^8-c_I\sum_{J<K}\delta^{IJK}s_Js_K{\cal N}_{\phi}^{32},
 \eea
where the functions $D_I, D$ are
$$
 D_I=c_I^2-\fr{H(y,x)}{H(x,y)}s_I^2,\quad D=D_1D_2D_3.
$$
Using them, we obtain the metric of the three-charge doubly rotating black ring as
 \bea
 ds_5^2=&-&D^{-2/3}\frac{H(y,x)}{H(x,y)}(dt+\Omega')^2+D^{1/3}\Bigl(- \frac{F(x,y)}{H(y,x)} d\phi^2 -
2\frac{J(x,y)}{H(y,x)} d\phi d\psi+\nn\\&+& \frac{F(y,x)}{H(y,x)} d\psi^2 +
\frac{2 k^2 H(x,y)}{(x-y)^2 (1-\nu)^2} \left[ \frac{dx^2}{G(x)} -
\frac{dy^2}{G(y)}\right]\Bigr),\label{chPomSen}
 \eea
with the rotation form
\be
\Omega'=\Omega_{\psi}'d\psi+\Omega_{\phi}'d\phi,\quad
\Omega_{\psi}'=\frac{\lambda_{78}'}{\lambda_{88}'}=c\Omega_{\psi}-s\tilde\Omega_{\phi},\quad
\Omega_{\phi}'=a_{\phi}^8{}'+a_{\phi}^7{}'\Omega_{\psi}'=c\Omega_{\phi}-s\tilde\Omega_{\psi}.
\ee
The corresponding vector potentials one-forms are given by
\bea
 \hat A^I{}'&=&\Bigl[s_Ic_I2\lambda(1-\nu)(x-y)(1-\nu
xy)dt+\Bigl(H(x,y)\tilde\Omega_{\phi}c_I\sum_{J<K}\delta^{IJK}s_Js_K-H(y,x)\Omega_{\psi}s_I\sum_{J<K}\delta^{IJK}c_Jc_K\Bigr)d\psi+\nn\\
 &+&\Bigl(H(x,y)\tilde\Omega_{\psi}c_I\sum_{J<K}\delta^{IJK}s_Js_K-H(y,x)\Omega_{\phi}s_I\sum_{J<K}\delta^{IJK}c_Jc_K\Bigr)d\phi\Bigr]\frac{1}{D_I H(x,y)}.\label{A_I}
 \eea
 Finally, the scalar field can be read off from the corresponding eleven-dimensional three-charge black tube solution
\be
ds_{11}^2=D^{1/3}\Bigl(\frac{dz_1^2+dz_2^2}{D_1}+\frac{dz_3^2+
dz_4^2}{D_2}+\frac{dz_5^2+dz_6^2}{D_3}\Bigr) + ds_5^2. \ee
This five-dimensional part of the solution turns into that obtained
in \cite{bccgsw} via identification of the vector fields $\hat
A^1=\hat A^2=\hat A^3=\frac{1}{\sqrt3}A$ and the charge parameters
$c_1=c_2=c_3=c$, $s_1=s_2=s_3=s$. Like in this case, the metric
(\ref{chPomSen}) and the fields (\ref{A_I}) contain Dirac-Misner
strings which arise if the orbits of $\partial_{\psi}$ do not close
off at $x=\pm1$. This is the case since both the
$\Omega'_{\psi}(1,y)$ and the $A_{\psi}^I{}'(1,y)$ are proportional
to $\tilde\Omega_{\phi}(1,y)$, which does not vanish. However, now
one can remove the Dirac-Misner string turning off  any two of three
charges.   Indeed,
setting, e.g., $c_1=1=c_2$, $s_1=0=s_2$ and $c_3=c$, $s_3=s$ we will
have the five-dimensional metric \be ds_{5}^2=
-D^{-2/3}\frac{H(y,x)}{H(x,y)}(dt+c\Omega)^2
+D^{1/3}\Bigl(\frac{F(y,x)}{H(y,x)} d\psi^2- \frac{F(x,y)}{H(y,x)}
d\phi^2 - 2\frac{J(x,y)}{H(y,x)} d\phi d\psi+  \frac{2 k^2
H(x,y)}{(x-y)^2 (1-\nu)^2} \left[ \frac{dx^2}{G(x)} -
\frac{dy^2}{G(y)}\right]\Bigr), \ee where $\Omega$ now is the
Pomeranski-Senkov original function and the scalar and the
non-trivial vector fields being \be
 X^1{}'=X^2{}'=D^{1/3},\ X^3{}'=D^{-2/3},\quad \hat A^1{}'=0=\hat A^2{}',\quad
 \hat A^3{}'=\frac{s}{D H(x,y)}\Bigl[2c\lambda(1-\nu)(x-y)(1-\nu
xy)dt-H(y,x)\Omega\Bigr],
\ee
with
$$
D=1+s^2\frac{2\lambda(1-\nu)(x-y)(1-\nu xy)}{H(x,y)}.
$$
This black ring is free from singularities. In this case the
solution corresponds to the particular case of the two-charge black
ring obtained by  Hoskisson \cite{hos}, if one of the charges is set
to zero.

Physical parameters of the solution are:
$$
 M=\frac{\pi k^2\lambda(3+2s^2)}{G_5(1-\lambda+\nu)},\quad
J_\phi{} = \frac{2 c k^3 \pi \lambda (1 + \lambda - 6 \nu + \lambda
\nu + \nu^2) \sqrt{(1 + \nu)^2 - \lambda^2}}{G_5 (1 - \nu)^2 (1 -
\lambda + \nu)^2},
$$
$$
J_\psi{} = \frac{4c k^3 \pi \lambda \sqrt{\nu} \sqrt{(1 + \nu)^2 -
\lambda^2}}{G_5 (1 - \nu)^2 (1 - \lambda + \nu)},\quad
Q=-\frac{2\lambda cs\pi k^2}{G_5(1-\lambda+\nu)}.
$$

The reason why this solution is so simple is the following.
One may observe that once two vector fields are set to zero while the scalar fields satisfy the condition $X_1=X_2$ the five-dimensional theory reduces to
 \be
 {\cal L}_5=R_5\star_5
 1-\frac12d\varphi\wedge\star_5d\varphi-\frac12\e^{-\frac{4\varphi}{\sqrt6}}\hat F^3\wedge\star_5
 \hat F^3,\quad \hat F^3=d\hat A^3,\quad \varphi=\frac{\sqrt6}{2}\ln X^3.\label{L_5}
 \ee
This Lagrangian  turns out to be the  Kaluza-Klein reduction of the $6D$ Einstein-Hilbert Lagrangian ${\cal
L}_6=R_6\star_6 1$. Therefore our one-charge non-singular solution can be uplifted to the following $6D$ vacuum solution
\be
 ds_6^2=\e^{\frac{\varphi}{\sqrt6}}ds_5^2+\e^{-\frac{3\varphi}{\sqrt6}}\left(dx^6+\frac{s}{D H(x,y)}\Bigl[2c\lambda(1-\nu)(x-y)(1-\nu
xy)dt-H(y,x)\Omega\Bigr]\right)^2.
\ee

 \section{Conclusion}
In this paper we have applied the $SO(4,4)$ generating technique to
get a charged version of the PS doubly rotating black ring. Our
general solution describes three charge black ring (or three charge
black tube) with two rotation parameters, being a six-parametric
solution. Unfortunately for arbitrary values of charges the solution
is plagued with Dirac-Misner strings. But setting zero two of the
three charges one gets a particularly simple charged doubly rotating
black ring. In view of its simplicity, this solution can be used for
further applications of generating transformations. Our present
reduction scheme has only one class of transformations preserving
asymptotic flatness, and these generate the electric charges.
Further progress in generating black rings amounts to generating the
dipole charges. Tentatively, this may be achieved using  dimensional
reduction to three dimensions with respect to a linear combination
of two rotation angles.

 We have also given the matrix representation of the coset $SO(4,4)/(SO(2,2)\times SO(2,2))$ in a more concise form than in the original formulation of our technique \cite{gtu1_3}.

  \begin{acknowledgments}
The authors are grateful to Gerard Cl\'ement
helpful discussions. The work was supported by the RFBR grant
08-02-01398-a. NS acknowledges the support by Dynasty foundation.
 \end{acknowledgments}


\end{document}